\documentclass[aps,prl,superscriptaddress,reprint,floatfix]{revtex4-1}
\usepackage{float}
\usepackage{amsmath}
\usepackage{graphicx}
\usepackage{longtable}
\usepackage{colortbl}
\usepackage[english]{babel}
\usepackage{array}
\usepackage{multirow}
\usepackage{tabularx}
\usepackage{booktabs}

\begin{document}

\title{Importance of Nuclear Quantum Effects on the Hydration of Chloride Ion}
%\title{Convergence of Theory and Experiment Reveals the Surprising Role of Quantum Effects in Chloride Solvation}
\author{Jianhang Xu}
\affiliation{Department of Physics, Temple University, Philadelphia, PA 19122}
\author{Zhaoru Sun}
\affiliation{Department of Physics, Temple University, Philadelphia, PA 19122}
\author{Chunyi Zhang}
\affiliation{Department of Physics, Temple University, Philadelphia, PA 19122}
\author{Mark DelloStritto}
\affiliation{Institute for Computational Molecular Science, Temple University, Philadelphia, PA 19122}
\author{Michael L. Klein}
\affiliation{Department of Physics, Temple University, Philadelphia, PA 19122}
\affiliation{Institute for Computational Molecular Science, Temple University, Philadelphia, PA 19122}
\affiliation{Department of Chemistry, Temple University, Philadelphia, PA 19122}
\author{Deyu Lu}
\affiliation{Brookhaven National Laboratory Center for Functional Nanomaterials, Upton, NY 11973}
\author{Xifan Wu}
\affiliation{Department of Physics, Temple University, Philadelphia, PA 19122}
\affiliation{Institute for Computational Molecular Science, Temple University, Philadelphia, PA 19122}

\date{\today}

\begin{abstract}
Path-integral \textit{ab initio} molecular dynamics (PI-AIMD) calculations have been employed to probe the nature of chloride ion solvation in aqueous solution. 
Nuclear quantum effects (NQEs) are shown to weaken hydrogen bonding between the chloride anion and the solvation shell of water molecules. 
As a consequence, the disruptive effect of the anion on the solvent water structure is significantly reduced compared to what is found in the absence of NQEs. 
The chloride hydration structure obtained from PI-AIMD agrees well with information extracted from neutron scattering data. 
In particular, the observed satellite peak in the hydrogen-chloride-hydrogen triple angular distribution serves as a clear signature of NQEs. 
The present results suggest that NQEs are likely to play a crucial role in determining the structure of saline solutions.
 \end{abstract}

\maketitle

%\section{INTRODUCTION}

%\subsection{Importance of study Cl}
\par 
Hydrated chloride ions (Cl$^-$) are ubiquitous in nature. 
They are an essential component in the electrolytes of living systems \cite{jentsch_molecular_2002}.
Also, Cl$^-$ is a member of the Hofmeister series of ions \cite{Hofmeister_zur_1888}, with important effects on protein solubility and folding. 
Moreover, chloride ion channels are a diverse group of anion-selective channels involved in the excitability of skeletal, cardiac, and smooth muscle cells \cite{Hofmeister_zur_1888}. 
These important biochemical and physiological roles all involve Cl$^-$ in an aqueous environment. 
Not surprisingly, the hydration structure of the chloride anion, Cl$^-$(aq), and its impact on the hydrogen (H)-bonding network of water have been the subject of intense scientific research for many decades \cite{cummings_chloride_1980,ohtaki_structure_1993,leberman_effect_1995,knipping_experiments_2000,ghosal_electron_2005,marcus_effect_2009,piatkowski_extreme_2014}.

%\subsection{current study and issues}
\par
The arrangement of water molecules around a Cl$^-$ can be probed by scattering experiments \cite{symons_neutron_2001,mancinelli_perturbation_2007,mancinelli_hydration_2007}, 
and the perturbed H-bond structure is inferable from spectroscopic measurements\cite{dang_molecular_2006,smith_effects_2007,kulik_local_2010,migliorati_unraveling_2014}. 
But most experiments typically yield only time-averaged structural information. 
At the molecular level, the solvation structure of Cl$^-$ is constantly fluctuating on a sub-picosecond time scale.
In this regard, \textit{ab initio} molecular dynamics (AIMD) simulation \cite{car_unified_1985} has already proven to be a valuable theoretical tool. %\cite{kulik_local_2010,ge_linking_2013,zhang_communication:_2013,gaiduk_structural_2014,pham_structure_2016,gaiduk_local_2017}. 
In AIMD, the forces needed to propagate the dynamics are generated from the instantaneous ground state based on density functional theory (DFT) \cite{kohn_self-consistent_1965}. 
AIMD can directly model the fast exchange of water molecules within the anion’s hydration shell, as well as the H-bond fluctuations in the water solvent. 
AIMD simulations of chloride in solution, Cl$^-$(aq), have been carried out since the 1990s \cite{laasonen_ab_1997,tobias_surface_2001,heuft_density_2003,kulik_local_2010,ge_linking_2013,zhang_communication:_2013,bankura_hydration_2013,gaiduk_structural_2014,bankura_systematic_2015,pham_structure_2016,gaiduk_local_2017}. 
Consensus has been reached on the fact that the water structure in the first hydration shell is strongly distorted. 
The chloride anion, as a H-bond acceptor, is polarized in solution due to its large size \cite{perera_structure_1992,ohtaki_structure_1993,tobias_surface_2001,ge_linking_2013,migliorati_unraveling_2014,dellostritto_aqueous_2020}. 
Therefore, the distribution of water molecules in its first coordination shell is rather inhomogeneous \cite{ge_linking_2013}. 
Such a defect-like solvation pattern around Cl$^-$ is incompatible with the tetrahedral structure of water and disrupts the hydrogen bond network in the solution.
Beyond the first solvation shell, recent AIMD simulations \cite{gaiduk_local_2017} carried out using the Perdew–Burke-Ernzerhof (PBE) functional \cite{perdew_generalized_1996} at an elevated temperature of 400K found a well-structured second solvation shell for Cl$^-$ and weakened hydrogen bonds as far as the third solvation shell. 
These previous studies have provided important insights on Cl$^-$(aq), but some issues remain unresolved. 
For example, one might expect that the water structure in saltwater is noticeably different from that of pure water. 
However, an analysis based on neutron scattering data surprisingly suggested that the disruption of the water structure by solvated Cl$^-$ is negligible beyond the first shell \cite{mancinelli_perturbation_2007,mancinelli_hydration_2007}. 

%\subsection{important of NQE on modify H-bond network}
\par 
Rationalization of the neutron scattering data requires atomic details on the solvation structure of the Cl$^-$.
In order to tackle this problem quantitatively with AIMD simulations, one needs to employ an accurate exchange-correlation functional. 
Moreover, treatment of nuclear quantum effects (NQEs) associated with the system’s protons is not optional, but indispensable in order to produce a liquid water structure compatible with the experimental observation \cite{tuckerman_quantum_1997,morrone_nuclear_2008,ceriotti_nuclear_2013,sun_electron-hole_2018}. 
Notably, the role of NQEs varies significantly among different types of H-bonds \cite{li_quantum_2011}. 
This new twist elevates the level of complexity in the computations. 
Two distinct types of H-bonds exist simultaneously in Cl$^-$(aq), namely the water-water (W-W) and anion-water (A-W) H-bonds, respectively. 
The former tends to build an extended tetrahedral network \cite{chen_ab_2017}, while the latter tends to form a tight A-W cluster surrounded by additional solvent water molecules. 
The resulting Cl$^-$ hydration structure reflects a delicate balance between these two competing effects.
NQEs tilt the balance between these competing H-bonding forces, which in turn can lead to a different hydration structure than is modeled using classical nuclei.

%\subsection{outline of our study}
\par 
The present work is focused on probing the structure of Cl$^-$(aq) via Feynman path integral \cite{feynman_quantum_1965} \textit{ab initio} molecular dynamics (PI-AIMD) simulations, as well as traditional AIMD with classical nuclei. 
The nuclear potential energy surface is generated by employing the recently introduced SCAN functional \cite{sun_strongly_2015}. 
Surprisingly, NQEs tilt the balance between the competing W-W and A-W H-bonding and give rise to important changes to the anion’s hydration structure. 
Under the influence of NQEs, both types of H-bonds are weakened. 
However, the A-W H-bond is weakened to a greater extent than the corresponding W-W H-bond. 
As a result, water molecules in the first hydration shell are relatively less tightly bound by the anion, and thus more amenable to accommodating the water solvent structure. 
While the first hydration shell still disrupts the water structure, surprisingly the solvent partially recovers its tetrahedral order. 
When compared to results based on classical nuclei, the PI-AIMD simulation show that the influence of Cl$^-$ on the water structure, beyond the first shell, is much weaker and, importantly, the solvent H-bond network is seemingly rapidly restored to its bulk-like behavior. 
The PI-AIMD result shows excellent agreement with the experiments of Soper \textit{et al} \cite{mancinelli_hydration_2007,mancinelli_perturbation_2007}. 
In particular, the satellite peak of the H-Cl-H angular distribution derived from neutron scattering data only appears in the PI-AIMD simulations, and is absent in conventional AIMD. 
Thus, NQEs give important corrections to the computed Cl$^-$(aq) hydration structure, yielding more consistent results when compared to experiments. 
The present findings strongly suggest that NQEs should be included in future studies of the Hofmeister series.

%\section{METHODS}

%\section{Difference between Cl and H2O interaction}
\begin{figure}[t]
\centering
\includegraphics[width=0.99\columnwidth]{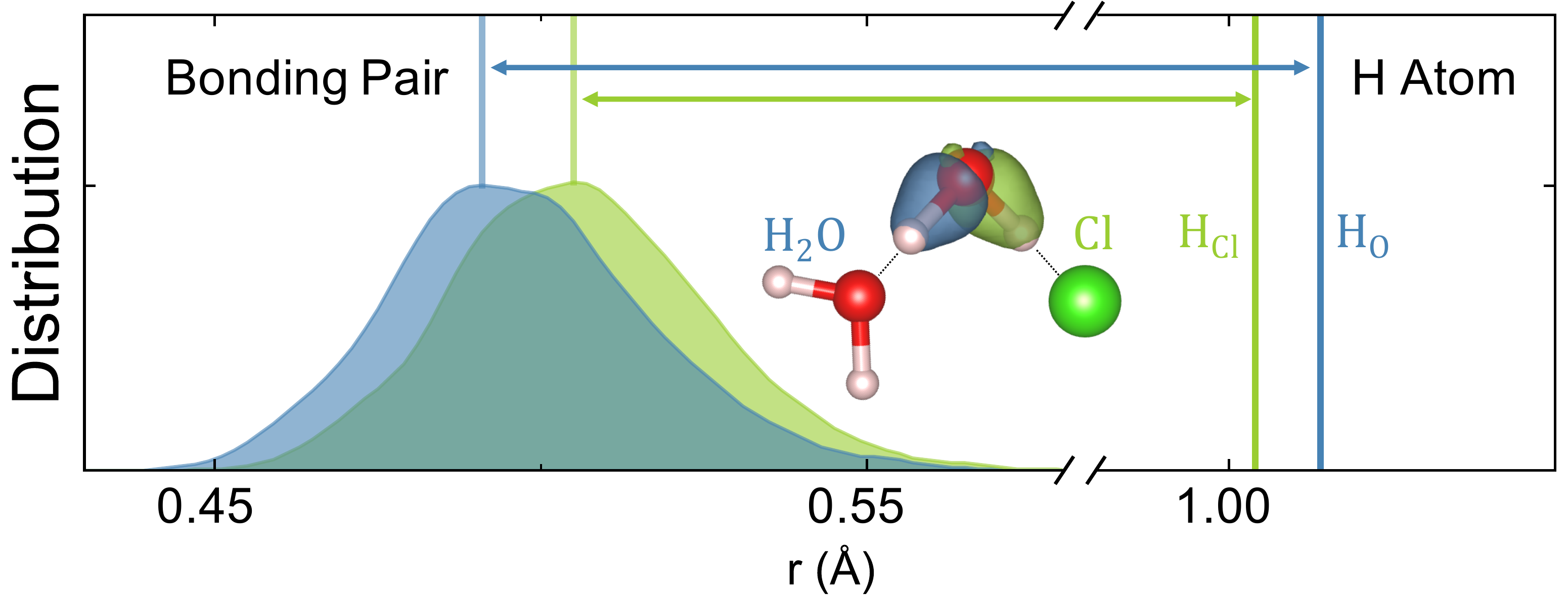}
\caption{
In the PI-AIMD trajectory, the distribution of the distances from the intramolecular oxygen to the MLWF centers for the bonding pair electrons; the vertical lines denote the average positions of two types of protons bonded to solvent water (blue) or to the Cl$^-$ anion (green), as schematically shown in the inset together with density isosurfaces of MLWF for the bonding pair electrons.}
\label{Wannier}
\end{figure}

\par
All AIMD and PI-AIMD calculations were performed in the canonical (NVT) ensemble at T=300K using a periodically replicated cubic box with edge length 12.42 \r{A}. 
One Cl$^-$ ion and 63 H$_2$O molecules were included in the 0.87 M Cl$^-$ aqueous solution. 
AIMD and PI-AIMD pure water simulations with 64 water molecules were performed as a comparison. 
All simulations employed the SCAN functional. 
Maximally localized Wannier Function (MLWF) \cite{marzari_maximally_1997,marzari_maximally_2012} centers were computed to study electronic properties. 
(Additional simulation details are provided in the Supplemental Materials.)

\par
As already mentioned, a water molecule in the first hydration shell of Cl$^-$ is subjected to competing forces provided by A-W and W-W H-bonds. 
Thus, one proton in the water molecule points towards the Cl$^-$ anion, while the other points to the lone pair electrons of a solvent water, as illustrated schematically in the inset of Fig. \ref{Wannier}.
Furthermore, this hydration shell water molecule is polarized by this special H-bonding configuration. 
Under its polarizing effect, the electropositive proton and the electronegative bonding pairs are separated further apart from each other generating a larger electric dipole in the condensed phase than that in water vapor \cite{badyal_electron_2000,eisenberg_structure_2005}. 
However, the abilities to polarize water are different for these two types of H-bonds as determined by the electronic structural properties. 
The A-W H-bond has a weaker bonding strength than that of W-W, as evidenced by the shorter distance between the bonding electron pairs and the proton in Fig. \ref{Wannier} \cite{chen_hydroxide_2018}. 
The relative weaker A-W bond also reduces the electric dipole of water molecules in the first hydration shell by $\sim$3\% compared to that in bulk water; 
an effect which has also been reported in the literature \cite{schmidt_water_2009,guardia_ion_2009}.

%\section{PTC}
\begin{figure}[t]
\centering
\includegraphics[width=0.90\columnwidth]{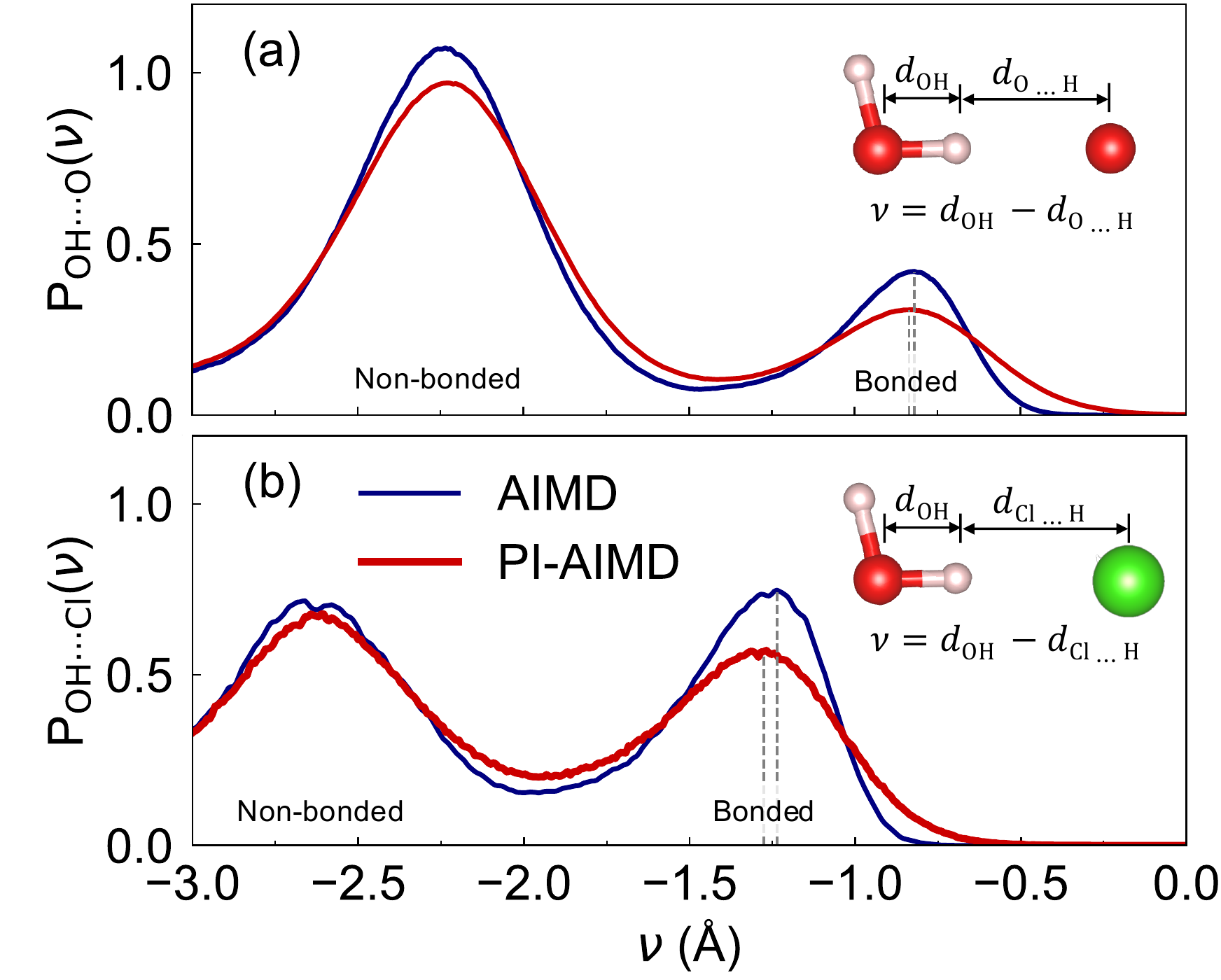}
\caption{
Spatial distributions of proton transfer coordinate $\nu$ for water molecules in Cl$^-$ (aq). 
(a) in the solvent (b) in anion’s first hydration shell in both AIMD (blue) and PI-AIMD (red) trajectories of Cl$^-$(aq). 
$\nu$ is defined as $\nu_X=d(\text{O-H})-d(X\text{-H})$, where the $X$ atom denotes either an oxygen atom or a chloride ion.
}
\label{ptc}
\end{figure}

\begin{figure*}[t]
\centering
\includegraphics[width=0.99\textwidth]{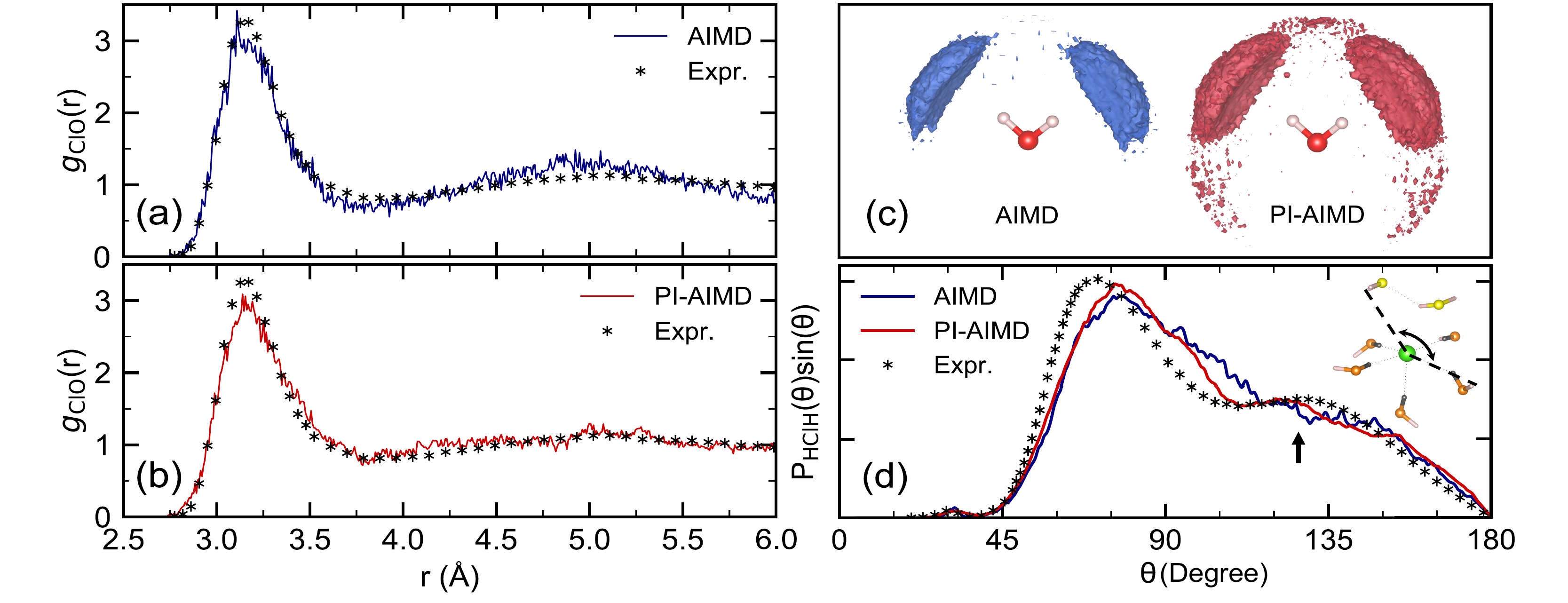}
\caption{
Cl-O Radial Distribution Functions in Cl$^-$ (aq) obtained from (a) AIMD (blue) and (b) PI-AIMD (red) simulations, with the neutron diffraction experimental result \cite{mancinelli_hydration_2007} (black) presented for comparison.
(c) Isosurface of probability of finding a Cl$^-$ ion first neighbor of a water molecule in AIMD (blue) and PI-AIMD (red) simulations. 
The contrast level of the isosurfaces is set to 0.70. 
(d) Probability distribution of H-Cl-H angles in the first hydration shell of Cl$^-$ solution obtained from AIMD (blue) simulation, PI-AIMD (red) simulation, and neutron diffraction experiment \cite{mancinelli_hydration_2007} (black). 
The inset shows polarized Cl$^-$ hydration structure, where water molecules in the first hydration shell are classified by bonded (orange) and non-bonded (yellow) to the ion.
}
\label{ClO}
\end{figure*}

\par 
Besides its impact on the electronic structure, H-bonding also affects the proton position \cite{tuckerman_quantum_1997}. 
The above effect is explored via the proton transfer coordinate ($\nu$) \cite{ceriotti_nuclear_2013,sun_electron-hole_2018}. The resulting distribution functions, P$_{\text{OH}\dotsi\text{O}}$($\nu$) and P$_{\text{OH}\dotsi\text{Cl}}$($\nu$) are shown in Fig. \ref{ptc}(a) and (b) for the water molecules in the bulk solvent as well as in the first hydration shell of Cl$^-$, respectively. 
In general, two distinct features can be identified in P($\nu$). 
The feature at more negative $\nu$ (around -2.2 \r{A} in P$_{\text{OH}\dotsi\text{O}}$($\nu$) and -2.7 \r{A} in P$_{\text{OH}\dotsi\text{Cl}}$($\nu$)) is contributed by the non-bonded hydrogen. 
Whereas, the other feature  (around -0.8 \r{A} in P$_{\text{OH}\dotsi\text{O}}$($\nu$) and -1.2 \r{A} in P$_{\text{OH}\dotsi\text{Cl}}$($\nu$), denoted as P$_{\text{OH}\dotsi\text{O}}^{\text{B}}$($\nu$) and P$_{\text{OH}\dotsi\text{Cl}}^{\text{B}}$($\nu$)) originates from the bonded hydrogen atoms via the W-W or the A-W H-bonds in Fig. \ref{ptc}(a) and (b), respectively. 
The shorter distance in the bonded peaks is attributed to the fact that protons are more likely to approach the acceptors, e.g. an enhanced tendency of proton transfer \cite{ceriotti_nuclear_2013,chen_hydroxide_2018}, under the attractive H-bond force. 
As expected, the peak positions of P$_{\text{OH}\dotsi\text{O}}^{\text{B}}$($\nu$) and P$_{\text{OH}\dotsi\text{Cl}}^{\text{B}}$($\nu$)) are the equilibrium positions of protons determined by the average strength of H-bonds under thermal fluctuations. 
The different peak positions of P$_{\text{OH}\dotsi\text{O}}^{\text{B}}$($\nu$) and P$_{\text{OH}\dotsi\text{Cl}}^{\text{B}}$($\nu$) are mainly caused by the size difference between the Cl$^-$ anion and oxygen atom \cite{shannon_revised_1976}.

\par 
Both types of H-bonds undergo notable changes when the protons are treated with NQEs in PI-AIMD simulations.
On the one hand, the zero-point motion significantly expands the region that protons are able to explore on the potential energy surface, which is inaccessible to classical nuclei. 
Therefore, both P$_{\text{OH}\dotsi\text{O}}$($\nu$) and P$_{\text{OH}\dotsi\text{Cl}}$($\nu$) shows a broader distribution in PI-AIMD trajectories. 
On the other hand, the centers of P$_{\text{OH}\dotsi\text{O}}^{\text{B}}$($\nu$) and P$_{\text{OH}\dotsi\text{Cl}}^{\text{B}}$($\nu$) move further away from its acceptors, which suggests that both H-bonds are weakened in PI-AIMD. 
The fact that H-bonding is weakened by NQEs has been recognized recently in pure water, which yields an important refinement to theoretical descriptions of water structure \cite{ko_isotope_2019,cheng_ab_2019}.
More importantly, H-bonding strength varies among different types of H-bonds. 
The protons are more delocalized by NQEs. 
However, while delocalization of the proton along the stretching direction facilitates H-bond formation, delocalization due to proton libration tends to weaken the H-bond. 
The result represents a delicate balance of the aforementioned opposing NQEs. 
A rule of thumb proposed by Michaelides \textit{et al.} states that the relatively weak H-bond will become even weaker by the NQEs and \textit{vice versa} \cite{li_quantum_2011}. 
Indeed, a close inspection reveals that the peak position of P$_{\text{OH}\dotsi\text{O}}^{\text{B}}$($\nu$) has a much larger shift moving away from acceptors than that of P$_{\text{OH}\dotsi\text{Cl}}^{\text{B}}$($\nu$).
Importantly, this means that the A-W H-bond becomes even weaker than the W-W H-bond under NQEs.

\begin{figure*}[t]
\centering
\includegraphics[width=0.99\textwidth]{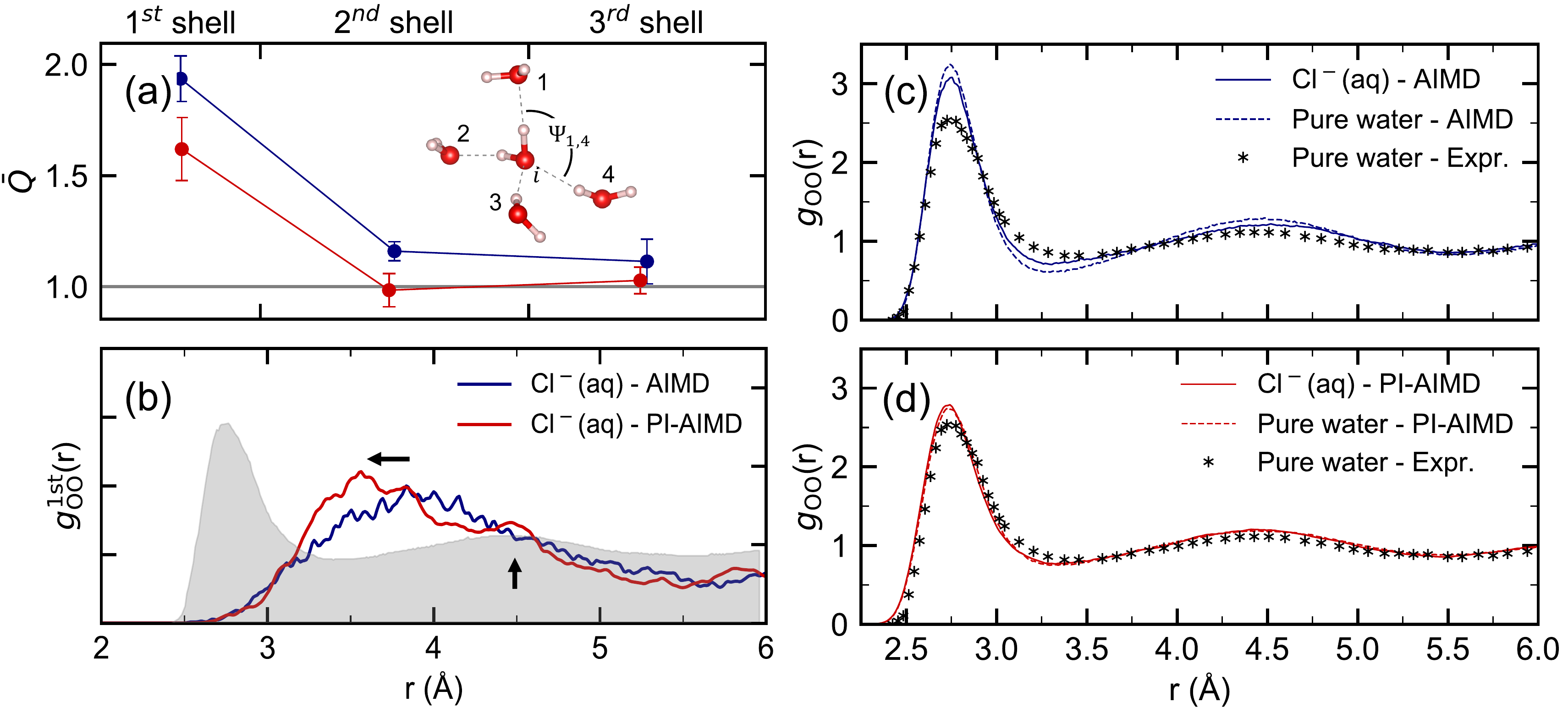}

\caption{
(a) Normalized tetrahedral structure order parameter $\bar{Q}$ decay as a function of the distance to Cl$^-$ ion of AIMD (blue) and PI-AIMD (red) simulations. 
For one water molecule $i$, $Q_i=\langle\sum_{j=1}^3\sum_{k=j+1}^4(cos\psi_{j,k}+\frac{1}{3})^2\rangle$, where $j$ and $k$ are the $j$th and $k$th nearest neighbor of water molecule $i$, and $\psi_{j,k}$ is the angle between molecule $i,j$ and $i,k$. 
$\bar{Q}$ defined by ${\langle Q_i \rangle_\text{solution}}/{\langle Q_i \rangle_\text{water}}$, is the averaged and normalized $Q$.  
(b) AIMD and PI-AIMD O-O RDFs within the first Cl$^-$ hydration shell of bonded water, along with regular O-O RDFs of bulk water. 
And O-O RDFs ($g_{_\text{OO}}$(r)) of (c) AIMD (solid blue) and (d) PI-AIMD (solid red) Cl$^-$ solution simulations, AIMD (dotted blue) and PI-AIMD (dotted red) pure water simulations, and diffraction experiment \cite{soper_quantum_2008} (black triangle).
}
\label{OO}
\end{figure*}

%\section{Structure changes first shell

\par 
Figure. \ref{ClO}(a) and (b), present the Cl-O radial distribution functions (RDFs) $g_{\text{ClO}}$(r) from the Cl$^-$(aq) trajectories for both AIMD and PI-AIMD.
For comparison, the experimental $g_{\text{ClO}}$(r) derived from neutron scattering is also shown.
Clearly, the Cl-O interaction is overly structured in the AIMD simulation, Fig. \ref{ClO}(a). 
The artificially strengthened Cl-O attraction with classical nuclei brings the first (at 3.14 \r{A}) and second hydration shells (at 4.91 \r{A}) closer to the anion, relative to the experimental peaks, at 3.16 \r{A} and 5.09 \r{A}. 
Moreover, the first minimum and second maximum are more prominent than those in experiment. 
In contrast, the more weakened A-W H-bond due to NQEs should loosen the anion’s hydration shell. 
Indeed, the Cl-O interaction is weakened in the PI-AIMD simulation, as shown in Fig. \ref{ClO}(b).
Consistently, the center position of both first (at 3.16 \r{A}) and second hydration shells (at 5.05 \r{A}) increases and yields better agreement with the experimental values. 
At the same time, the overall $g_{\text{ClO}}$(r) RDF from PI-AIMD becomes much less structured, which then shows quantitative agreement with the experiment.

\par 
Because the Cl$^-$ anion is polarized in solution, the bonded water molecules tend to preferentially populate one side of the anion while leaving the other half-space relatively empty, with sporadic residence of non-bonded molecules (further details about hydration pattern is shown in the Supplemental Materials). 
To accept a H-bond, Cl$^-$ is located close to a proton along the bonding direction. 
The AIMD trajectory shows that Cl$^-$ is distributed within a narrow region with a double \textit{dome-like} shape as illustrated in Fig. \ref{ClO}(c). 
The overall solvation cage, composed of both bonded and non-bonded water molecules in the first hydration shell, can be described by a polyhedron with n $\sim$7 vertices, as illustrated schematically in the insert of Fig. \ref{ClO}(d). 
Consistent with the polyhedral geometry under thermal fluctuations, the H-Cl-H triple angular distribution P$_\text{HClH}$($\theta$) as plotted in \ref{ClO}(d) is centered at $\sim$70$^\circ$in the AIMD simulation, which is in qualitative agreement with experiment.
However, a second broader peak that appears clearly in experiment around 120$^\circ\sim$ 135$^\circ$ is absent in the AIMD simulation.

\par 
In the PI-AIMD trajectory, the solvation cage changes its geometry accordingly as a result of NQEs. 
Due to the weakened A-W H-bond, the average distance between the chloride ion and its bonded water molecules is slightly increased from 3.235 \r{A} to 3.249 \r{A}.
At the same time, the \textit{dome-like} distribution of Cl$^-$ spreads out over a larger area due to quantum fluctuations, as shown in Fig. \ref{ClO}(c). 
Again, because of the weaker A-W bonding, the Cl$^-$ can no longer bond as many water molecules as it does in the AIMD trajectory. 
As a consequence, the population of non-bonded water molecules largely increases by $\sim$50\% from $\sim$1.6 in AIMD to $\sim$2.4 in PI-AIMD (additional detail about changes of coordination number is discussed in the Supplemental Materials). 
The increased number of non-bonded waters in the first hydration shell can be further confirmed by the significantly increased distribution of Cl$^-$ in the region around the oxygen.
With more vertices occupied by non-bonded water molecules, the solvation cage predicted by PI-AIMD has geometric characteristics different from that of AIMD.
Because the bonded and non-bonded water molecules are located on opposite sides of Cl$^-$, the triplet angular distribution P$_\text{HClH}$($\theta$) that involves a non-bonded water molecule mostly contributes to an obtuse angle as demonstrated in the insert of Fig. \ref{ClO}(d). 
As a result, the second broad peak in P$_\text{HClH}$($\theta$) centered around 120$^\circ\sim$ 135$^\circ$ emerges in the PI-AIMD simulation, as seen in Fig. \ref{ClO}(d); 
a finding which is in excellent agreement with experiment.

%\section{Structure changes overall}

\par 
The presence of Cl$^-$ disrupts the H-bond network, and distortions are expected around the solvated ion. 
Fig. \ref{OO}(a) presents the tetrahedral structure order parameters, $Q$ \cite{chau_new_1998} of water solvent in different hydration shells, as computed from both AIMD and PI-AIMD trajectories. 
Not surprisingly, the most abrupt distortion takes place in the first hydration shell because the solvation cage polyhedron is intrinsically different from a tetrahedron. 
Consistently, the first peak of $g_\text{OO}^\text{1st}$(r), the O-O RDF computed by only water molecules in the first hydration shell, is also drastically different from bulk water as shown in \ref{OO}(b). 
Away from the solvated anion, the degree of distortion on H-bond network decays, and the tetrahedral water structure gradually recovers to its bulk value in the second hydration shell and beyond as shown in Fig. \ref{OO}(a). 
In the AIMD simulation, the remaining distortion is non-negligible for solvent structure in the second and third hydration shells. 
As a result, the overall$g_\text{OO}$(r) in the AIMD trajectory is softened as compared to that in bulk water modeled by AIMD. 
This effect has been attributed to a long-range structural disturbance on the underlying H-bond network by the chloride anion \cite{gaiduk_structural_2014,gaiduk_local_2017}.

\par 
Interestingly, the inclusion of NQEs in PI-AIMD trajectory lessens the impact of anion on the underlying H-bond network. 
A water molecule in Cl$^-$(aq) is attracted by two competing forces from the Cl$^-$ anion and the solvent water. 
The aforementioned more weakened A-W H-bond makes it easier for the water molecules to attract each other and restore the tetrahedral liquid structure. 
Indeed, the facilitated water structure can already be seen in the first hydration shell. 
As shown in Fig. \ref{OO}(b), the H-bond network is still largely disrupted in the first coordination shell. 
However, it becomes more structured; 
this can be seen by the decreased distance of the first peak, as well as the appearance of a second peak around 4.5 \r{A}, which coincides with the peak position of the second hydration shell of $g_\text{OO}$(r) in bulk water. 
Clearly, this indicates a partial recovery of the solvent water structure. 
By the same token, the normalized structural order parameter $\bar{Q}$ recovers more quickly its bulk value in the PI-AIMD simulation, as seen in Fig. \ref{OO}(a). 
The resulting overall $g_\text{OO}$(r) of the solvent in PI-AIMD is also very close to that in bulk water in Fig. \ref{OO}(d). 
The relatively small impact on the water structure caused by the Cl$^-$ anion, as modeled by including NQEs, agrees well with the conclusions of Soper \textit{et al}. based on analysis of their neutron scattering data \cite{mancinelli_hydration_2007} and by Funkner \textit{et al}. from terahertz absorption spectroscopy \cite{funkner_watching_2012}. 
Notably, inclusion of NQEs not only gives a more accurate description of Cl$^-$(aq) but also play an essential role in describing the bulk water structure more accurately. 
The computed $g_\text{OO}$(r) via PI-AIMD greatly helps to improve the agreement between theory and experiment.

%\section{CONCLUSION}

\par
In conclusion, NQEs have a surprisingly large influence on the hydration structure of Cl$^-$(aq). 
Specifically, the interaction between water and Cl$^-$ is weakened, so that the anion’s disruptive effect on the solvent H-bond network of solvent water is reduced.
The predicted hydration properties computed via PI-AIMD agree well with experiments. 
In particular, the emergence of the satellite peak in the H-Cl-H triangular distribution function in the PI-AIMD trajectory is a clear signature of NQEs.
The present results highlight the important role played by NQEs in ionic solutions involving the Hofmeister series. 
Complementary studies of NQEs on hydration of cations, such as Na$^+$(aq) and K$^+$(aq), should be interesting. 
Unlike the Cl$^-$ anion, these metal cations are not H-bonded to water in solution, and only the underlying water solvent will be affected by NQEs. 
Therefore, distinct corrections by NQEs are to be expected.

\begin{acknowledgments}
This work was primarily supported by the Computational Chemical Center: Chemistry in Solution and at Interfaces funded by the DoE under Award No. DE-SC0019394 (X.W. and M.L.K). 
This research used resources of the Center for Functional Nanomaterials, which is a U.S. DOE Office of Science Facility, at Brookhaven National Laboratory under Contract No. DE-SC0012704 (D.L.). 
It is partially supported as part of the Center for the Computational Design of Functional Layered Materials, an Energy Frontier Research Center funded by the U.S. Department of Energy, Office of Science, Basic Energy Sciences under Award No. DE-SC0012575 (J.X.). 
This work is also partially supported by the National Science Foundation through Grant No. DMR-1552287 (C.Z.). 
The computational work used resources of the National Energy Research Scientific Computing Center (NERSC), a U.S. Department of Energy Office of Science User Facility operated under Contract No. DE-AC02-05CH11231.
And this research includes calculations carried out on Temple University’s HPC resources and thus was supported in part by the National Science Foundation through major research instrumentation grant number 1625061 and by the US Army Research Laboratory under contract number W911NF-16-2-0189.
\end{acknowledgments}

\bibliography{cl}
\end{document}